\documentclass[twocolumns, abstract]{aa} %
\usepackage{graphicx}
\usepackage{natbib}
\bibpunct{(}{)}{;}{a}{}{,}
\usepackage{syntonly}
\usepackage[fleqn]{mathtools}
\usepackage{amssymb}
\usepackage{amsmath}
\usepackage{url}

\usepackage{xspace}

\usepackage{layouts}
\usepackage{color}
    \definecolor{Blue}{rgb}{0.0,0.0,1.0}
    \definecolor{Red}{rgb}{1.0,0.0,0.0}
    \definecolor{Green}{rgb}{0.0,1.0,0.0}

\newcommand{\gy}{\textsc{Gyoto}\xspace}

\newcommand{\be}{\begin{equation}}
\newcommand{\ee}{\end{equation}}
\newcommand{\bea}{\begin{eqnarray}}
\newcommand{\eea}{\end{eqnarray}}
\newcommand{\nn}{\nonumber}
\newcommand{\pp}{\varphi}
\newcommand{\dd}{\mathrm{d}}

\graphicspath{{Figure/}}
\begin{document}
\title{Multi-wavelength torus-jet model for Sgr~A*}

%
\author{     F. H. Vincent\inst{1}
\and         M. A. Abramowicz\inst{2,3,4}  
\and         A. A. Zdziarski\inst{2}
\and         M. Wielgus\inst{5,6} 
\and         \\T. Paumard\inst{1}
\and         G. Perrin\inst{1}
\and         O. Straub\inst{7,1} }
\institute{
         LESIA, Observatoire de Paris, PSL Research University, CNRS, Sorbonne Universit\'es, UPMC Univ. Paris 06, Univ. Paris Diderot, Sorbonne Paris Cit\'e, 5 place Jules Janssen, 92195 Meudon, France
             \\ \email{frederic.vincent@obspm.fr}
             \and              Nicolaus Copernicus Astronomical Center, Polish Academy of Sciences, Bartycka 18, PL-00-716 Warszawa, Poland
             \and              Physics Department, Gothenburg University, 412-96 Gothenborg, Sweden
             \and              Physics Department, Silesian University of Opava, Czech Republic
             \and              Harvard-Smithsonian Center for Astrophysics, 60 Garden St., Cambridge, MA 02138, USA
             \and              Black Hole Initiative at Harvard University, 20 Garden St., Cambridge,
             MA 02138, USA
             \and              Max Planck Institute for Extraterrestrial Physics, Giessenbachstr. 1, 85748 Garching, Germany
%
}
   \date{Received ; accepted }
\abstract
{The properties of the accretion/ejection flow surrounding the supermassive
  central black hole of the Galaxy, Sgr~A*, will be scrutinized by the
  new-generation instrument GRAVITY and the Event Horizon Telescope (EHT).
  Developing fast, robust, and simple models of such flows is thus important
and very timely.}
{We want to model the quiescent emission of Sgr~A* from radio
  to mid-infrared, by considering
  a magnetized compact torus and an extended jet.
  We compare model spectra and images to the multi-wavelength
  observable constraints available to date.
  }
   { We use a simple analytic description for the geometry of the torus and jet.
     We model their emission respectively by thermal synchrotron
     and $\kappa$-distribution synchrotron. We use relativistic ray tracing to
     compute simulated spectra and images, restricting our analysis
     to the Schwarzschild (zero spin) case. A best-fit is found by adjusting
     the simulated spectra to the latest observed data, and we check
     the consistency of our spectral best fits with the radio-image sizes,
     and infrared spectral index constraints.
     We use the open-source \texttt{eht-imaging} library to generate
   EHT-reconstructed images.}
 {We find perfect spectral fit ($\chi^2_\mathrm{red}\approx1$) both for
   nearly face-on and nearly edge-on views. These best fits give
   parameters values very close to that found by the most recent
   numerical simulations, which are much more complex than our model.
   The intrinsic radio size of Sgr~A* is found to be in reasonable agreement
   with the centimetric observed constraints. Our best-fit infrared spectral
   index is in perfect agreement with the latest constraints.
   Our emission region at $1.3$~mm, although larger than the
   Doeleman et al. (2008) Gaussian best-fit,
   does contain bright features at the $\lesssim 40\,\mu$as scale.
   EHT-reconstructed images show that torus/jet-specific features
   persist after the reconstruction procedure, and that these features
 are sensitive to inclination.}
 {The main interest of our model is to give a simple and fast model
   of the quiescent state of Sgr A*, which gives extremely similar
   results as compared to state-of-the-art numerical simulations.
   Our model is easy to use and we publish all the material necessary
   to reproduce our spectra and images, so that anyone interested
   can use our results rather straightforwardly.
   We hope that such a public tool can be useful in the context
   of the recent and near-future GRAVITY and EHT results.
   Our model can in particular be easily used to test
   alternative compact objects models,
   or alternative gravity theories.
   The main limitation of our model is that we do not yet treat the
   X-ray emission.}

\titlerunning{Torus-jet model for Sgr~A*}
\maketitle

%
%
\section{Introduction}
%
%
%

The supermassive black hole at the center of our Galaxy, Sgr~A*, is the best 
target for studying the vicinity of a black hole at high angular resolution.
With a mass of $\approx 4\times10^6\, M_\odot$ at a distance of $\approx 8$~kpc~\citep{ghez08,gillessen09b},
the angular size of this object~\citep[more precisely, of the black
hole shadow,][]{falcke00} is of $\approx 50~\mu$as, making it the
biggest black hole of the Universe on sky. It is thus of particular importance
to study the properties of the accretion flow surrounding this object, and compare
them with observable constraints.

The study of Sgr~A* is entering a new era with the advent
of $\approx 10-30~\mu$as-scale observations that are starting
to be delivered by the new-generation instruments GRAVITY~\citep{gravity17,gravity18b}
and the Event Horizon Telescope~\citep[EHT,][]{doeleman09}.
These instruments will, among other goals, allow to get a much more
precise understanding of the physics of the accreted gas close to the black hole.
In this perspective, modeling the electromagnetic radiation emerging from this accretion flow is important.

In this study, we focus on the quiescent emission of Sgr~A*,
when the source does not show outbursts,
or flares of radiation~\citep[see e.g.][and references therein]{doddseden11}.
For a complete review of Sgr~A* emission in the quiescent
and flaring states, see~\citet{genzel10}.
The quiescent radiation emitted in the region around 
Sgr~A* can be broadly divided as follows: 
\begin{itemize}
\item the radio spectrum ($1-100$~GHz, $3$~mm to $30$~cm)
is mainly due to non-thermal synchrotron~\citep{yuan03}, emitted far
from the black hole. The observed size is dominated by the scattering
effects~\citep[][]{bower06,falcke09}. The radio counterpart of Sgr~A*
has been studied since decades by means of Very Long Baseline
Interferometry~\citep[VLBI,][]{alberdi93,bower14};
\item the millimeter spectrum ($100$~GHz-$1$~THz, $0.3$~mm to $3$~mm) is
due to a mixture of thermal and non-thermal synchrotron~\citep{yuan03},
emitted very close to the black hole~\citep[inner few tens to hundreds
$\mu$as,][]{doeleman08}.
The EHT observes in this range at $1.3$~mm,
with future plans to enable observations at $0.86$~mm.
Advanced scattering mitigation
algorithms were recently developed to enable Sgr~A*
intrinsic imaging in the millimeter range \citep{johnson2016};
\item the infrared spectrum (between $\approx 1$ and $\approx 10\,\mu$m)
  is mainly due to non-thermal synchrotron
  radiation~\citep{yuan03,witzel18} emitted in the inner regions.
  GRAVITY observes in this range at $2.2\,\mu$m;
\item the X-ray spectrum ($2 - 10$~keV)
  is mainly due to thermal bremsstrahlung emitted
  at large scales in the central arcsecond~\citep{quataert02},
  with the addition of a
  small ($< 20$\%) contribution due to Compton
  scattering arising from the inner regions~\citep{wang13}.
\end{itemize}

Modeling Sgr~A* accretion/ejection flow in the aim of
accounting for part or all of these emission processes
has been a very intense area of research in the past decades.
Our Galactic center is an extreme case of a low-luminosity
galactic nucleus, radiating at $\approx 10^{-8}$ of the
Eddington level. As such, Sgr~A* is a prototype for the class
of hot accretion flows, for which most of the energy
is advected inwards and/or ejected as outflows,
rather than radiated away~\citep[see][for a review]{yuan14}.
It is practical to divide the publications into
those models that are analytical, and those using
numerical simulations (general relativistic magnetohydrodynamics
simulations, or GRMHD). Analytical studies can themselves
be divided into those dedicated to studying the emission
of geometrically thick hot disks, known as
radiatively inefficient accretion flows~\citep[RIAF,][]{narayan95nat,ozel00,yuan03,broderick16}, or ionized tori~\citep{rees82,straub12,vincent15},
and those studying the emission of a large-scale
jet~\citep{falcke93,falckemarkoff00,markoff01}. GRMHD simulations of Sgr~A* have
emerged a decade ago~\citep{moscibrodzka09,dexter10,shcher12,dibi12},
and soon became an extremely active field with the perspective
of the EHT observations~\citep[][citing only the most recent works,
see many other references therein]{ressler17,chael18,jimenez18,davelaar18}.

Although GRMHD models become more and more common with time, it is still
important to devote efforts to the development and use of  analytic models.
Indeed, analytic descriptions have the advantage of their simplicity:
a few carefully chosen parameters, with clear physical meaning,
describe only those physical effects that are
considered by the authors to be the relevant ones to account for
observable effects. Such a framework allows to get rid of the
big difficulty of discriminating the putative numerical artefacts
(typically linked to a particular initial or boundary condition)
that might impact the results of GRMHD simulations.
Moreover, analytic models are much faster, and thus well adapted to
scan parameter spaces, thus paving the way for future more demanding
GRMHD studies. In particular, analytical models are well adapted
for studying non-standard scenarios, like alternative compact
objects~\citep{vincent16a,vincent16b,lamy18}.

This article is devoted to expanding our past studies that aimed at accounting
for the emission of the surroundings of Sgr~A* with a simple
magnetized torus in the few tens of gravitational radii from the black hole.
This series of analyses started with~\citet{straub12,vincent15},
in which we showed that we could well fit the millimeter spectrum
of Sgr~A*, but were not able to account for the radio data, given
that larger-scale emission is needed for that. As a consequence, we
consider in this article the addition of a large-scale jet,
on top of the same magnetized torus as introduced in our past studies.
We note that a jet was already coupled to an advection-dominated
hot flow by~\citet{yuan02}, although not with ray tracing
as we do here, which does
not allow to predict the aspect of the millimetric images.
The existence of a jet at the Galactic center is supported
by the fact that such outflows have been detected in
other low-luminosity active galactic nuclei~\citep{falcke00b,bietenholz00}.
Moreover, numerical simulations demonstrate the natural link between
hot accretion flows and outflows~\citep{yuan14}. However, no
clear observational proof of the existence of a jet at the
Galactic center was yet obtained, and recent work by
\citet{issaoun19} favors either disk-dominated models
or face-on jet-dominated models as more likely to explain $3.5$~mm
spatially resolved emission from the Galactic center.
The presence of a jet at the Galactic center is thus
still an open question.

Our goal is to fit the radio to infrared spectrum of Sgr~A*.
We do not yet aim at accounting for the X-ray emission because this
would ask for still larger-scale simulations to take into account
the thermal bremsstrahlung that accounts for most of the quiescent X rays.
Also, our prime interest for later use of this model is the interpretation
of GRAVITY and EHT data, so that X rays are not our first target.

We insist on the fact that our model is fully open-source, and
readily available to be used by other authors without much effort.
Appendix~\ref{app:gyoto} gives the necessary and sufficient
information to be able to generate most of the numerical results
presented in this article.

Section~\ref{sec:model} presents the analytic torus+jet
model that we use, Section~\ref{sec:specim} gives our results
in terms of best-fit spectra and images,
Section~\ref{sec:ehtimaging} presents examples of synthetic
reconstructions of model images with a numerical
EHT array, and Section~\ref{sec:conc}
discusses our conclusions and perspectives.

\section{Torus-jet model}
\label{sec:model}
%
%

Throughout, the spacetime is assumed to be described by
the Kerr metric in Boyer-Lindquist $(t,r,\theta,\pp)$ coordinates,
describing a rotating black hole with mass $M$ and dimensionless
spin parameter $a$. In this article, we are not interested in the (small) effect
of the spin parameter, and will keep $a=0$ (Schwarzschild metric) throughout,
although our model is fully valid for any spin value.
The cylindrical radius is defined by $\rho = r\,\sin\theta$ (be careful that
$\rho$ is never a density in this article, it will always label the cylindrical radius).
The height is defined by $z = r\, \cos\theta$.


We describe below two structures that aim at being the sources of
the synchrotron radiation emitted around Sgr~A*:
\begin{itemize}
\item a magnetized torus which
gives a reasonable approximation of a
snapshot of a realistic accreting geometrically thick accretion flow.
This torus emits the thermal synchrotron responsible for the
millimeter peak of Sgr~A*;
\item a jet sheath, modeled as simply as possible to capture
only the crucial features of a realistic ejection flow. This
jet emits a mixed thermal/non-thermal synchrotron radiation that allows
to reproduce the radio spectrum of Sgr~A*, as well as
the mid- and far-infrared data.
\end{itemize}

We note that we do not consider the X-ray emission (which needs bremsstrahlung
and Comptonization) in this article.
This is postponed to a later study. This choice is dictated by our
prime interest in the infrared and millimeter instruments GRAVITY and the EHT.

\subsection{Torus model and its thermal synchrotron emission}

The torus structure exactly follows the description of~\citet{vincent15},
to which we refer for further details. In this article, we consider
a chaotic magnetic field \citep[i.e. isotropized, as compared to the
toroidal field of][]{komissarov06}, given that~\citet{vincent15} have
shown that the magnetic field directionality has no impact on the
spectral observables.

For completeness, we remind here the major features of this torus model.
We consider a circularly-rotating perfect fluid described by a constant
angular momentum $\ell = -u_\pp / u_t$, where $\mathbf{u}$ is the 4-velocity
of the fluid. The conservation of stress-energy leads to
\be
\frac{\nabla_\mu p }{p+\epsilon} = - \nabla_\mu \mathrm{ln}(-u_t)
\ee
where $p$ is the fluid pressure and $\epsilon$ is the fluid total energy density.
The constant-pressure surfaces are thus the same as the isocontours of the
potential $\mathcal{W} = \mathrm{ln}(-u_t)$. This leads to a toroidal shape
that is fully defined by the choice of $\ell$ together with the choice of the inner radius
$r_\mathrm{in}$ of the torus. The pressure is linked to enthalpy $h$ by means
of the polytropic equation of state $p = \kappa h^k$, where $k$ is the polytropic index.
The electron number density is then deduced from the enthalpy profile
and can be shown to be fully characterized by the potential $\mathcal{W}$
and a chosen averaged number density at the torus center $\langle n_e \rangle^\mathrm{T\;cen}$~\citep[see][for details]{vincent15}. Hereafter,
a superscript T labels a torus quantity, while a superscript J will labels a jet quantity.
The magnetic field is found by choosing the magnetization
parameter $\sigma$ (see Eq.~\ref{eq:sigma}). Finally, the electron temperature $T_e^\mathrm{T}$ varies as $\left( \langle n_e\rangle^\mathrm{T}\right)^{k-1}$,
with a scaling defined by the central temperature, $T_e^\mathrm{T\;cen}$, which is a free
parameter of the model.
{We refer the reader to Fig.~2 of~\citet{vincent15} and Fig.~2 of~\cite{straub12} for a description of the density and temperature profiles of the torus,
  and a comparison with the RIAF model. We stress that our torus is very compact
  and restricted to the inner $\approx 15$ gravitational units.}

We consider a population of thermal electrons at any point inside the
torus, with a number density distribution satisfying
\be
n_e^\mathrm{ther}(\gamma) = \frac{\langle n_e \rangle^\mathrm{T}}{\theta_e} \, \frac{\gamma (\gamma^2-1)^{1/2}}{K_2(1/\theta_e)} e^{-\gamma/\theta_e}
\ee
where $\langle n_e \rangle^\mathrm{T}$ is the energy-averaged
electron number density in the torus, $\gamma$ is the Lorentz factor of the electrons, $\theta_e = k T_e^\mathrm{T} / m_e c^2$
is the dimensionless electron temperature ($k$ is the Boltzmann constant, $c$ the velocity of light), and $K_2$ is a modified Bessel function of the second kind. We remind that the averaged number density $\langle n_e \rangle^\mathrm{T}$ and temperature $T_e^\mathrm{T}$ are analytically known
at any point of the torus.

The thermal synchrotron emission and absorption coefficients are different
from~\citet{vincent15}. Here, we consider the formula provided by~\citet{pandya16}, in their Eq.~31,
while~\cite{vincent15} consider the formula of~\citet{wardzinski00}.
This change is made for consistency with the jet emission, which is
taken from~\citet{pandya16}. 
The thermal synchrotron emission coefficient (in units of $\rm{erg\,s^{-1}\,cm^{-3}\,ster^{-1}\,Hz^{-1}}$) reads
\be
j_\nu^\mathrm{ther} = \frac{\langle n_e \rangle^\mathrm{T} \,e^2\, \nu_c}{c} \,e^{-X^{1/3}}\,\frac{\sqrt{2}\pi}{27} \, \sin\theta_B\,(X^{1/2} + 2^{11/12}X^{1/6})^2
\ee
where
$X = \nu / \nu_s$, $\nu_s = 2/9 \, \nu_c\,\theta_e^2\,\sin\theta_B$,
$\nu_c = e\,B / (2\pi\,m_e\,c)$ is the cyclotron frequency, with $e$
the electron charge, $m_e$ their mass, $B$ the magnetic field magnitude, and $\theta_B$ is the angle between the magnetic field direction and
the direction of photon emission (over which the emission is averaged). The absorption coefficient (in units of $\rm{cm}^{-1}$)
is found from Kirchhoff's law, $\alpha_\nu^\mathrm{ther} = j_\nu^\mathrm{ther} / B_\nu$,
where $B_\nu$ is the Planck blackbody function.

\subsection{Jet model and its synchrotron emission}

\begin{figure*}[htbp]
\centering
\includegraphics[width=0.3\hsize]{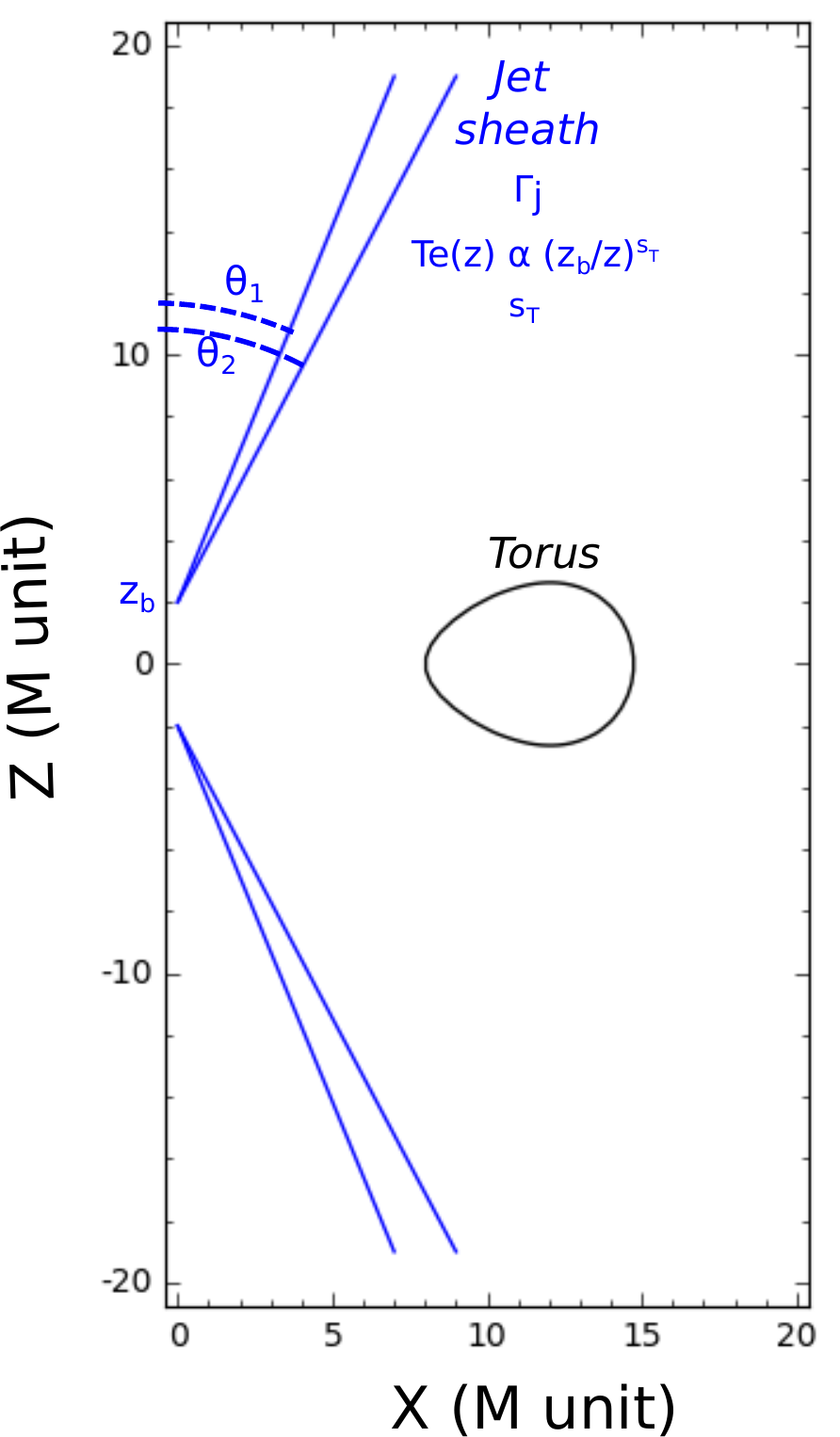}
\caption{Scheme of the torus-jet model. The jet is parametrized by 
the angles $\theta_1$ and $\theta_2$ that describe the angular opening
of the radiation-emitting sheath, by the base height $z_b$, the constant Lorentz
factor $\Gamma_j$, and the temperature power-law index $s_T$. The jet is symmetrical with respect to the equatorial plane, and axisymmetric.
} 
\label{fig:jetschema}
\end{figure*}

We want to keep our analytical jet model as simple as possible,
and close to the recent disk-jet 
GRMHD simulations obtained for Sgr~A* by~\citet{moscibrodzka13,davelaar18}.
These simulations show in particular that the radiation from the jet actually
comes from a narrow sheath, while most of the interior of the jet (the spine)
is empty of matter and thus does not contribute to the emission.
This finding was confirmed by~\citet{ressler17}.
We thus consider the model illustrated in Fig.~\ref{fig:jetschema}.
The emitting region is assumed to be defined by a thin layer in between
two conical surfaces defined by the angles $\theta_1$ and $\theta_2$,
and truncated at the jet base height $z_b$. The bulk Lorentz factor, as measured by the zero-angular-momentum observer (ZAMO\footnote{We remind that the ZAMO
is defined in Boyer-Lindquist coordinates by having zero angular momentum, $u_\pp=0$, at some fixed $r$ in the equatorial plane $\theta=\pi/2$. This fully fixes the ZAMO 4-velocity. In the Schwarzschild metric, such an observer is simply static. In the Kerr metric with non-zero spin, the ZAMO has a varying $\pp$ coordinate due to frame-dragging.}), is assumed
constant at $\Gamma_j$. The acceleration zone is thus discarded in this
simple model. 

Following~\citet{davelaar18} we consider 
a population of electrons, at any point inside the jet sheath, satisfying a $\kappa$-distribution
\be
n_e^\mathrm{\kappa-distrib}(\gamma) = N \,\gamma (\gamma^2 - 1)^{1/2} \left( 1+ \frac{\gamma-1}{\kappa \theta_e}\right)^{-(\kappa+1)}
\ee
where $N$ is a normalization factor depending on  the averaged electron number density in the jet, $\langle n_e \rangle^\mathrm{J}$, and the dimensionless temperature $\theta_e$ ($N$ is defined by $\int_\gamma n_e^\mathrm{\kappa-distrib}(\gamma) \dd\gamma=\langle n_e \rangle^\mathrm{J}$), $\gamma$ is the Lorentz factor of the electrons, and $\kappa$ is a parameter.
This distribution smoothly connects a thermal distribution for small
electron Lorentz factor, to a power-law distribution with power-law index $p = \kappa-1$ at high electron Lorentz factor. See Fig.~1 of~\citet{pandya16} for
an illustration of this distribution~\citep[note that the thermal/power-law transition
takes place close to the peak of the distribution; this is different from
the thermal+power-law-tail spectrum of the hard state of Cygnus X-1 as
discussed in][]{mcconnell02}.

The averaged electron number density varies with the altitude $z$. By the conservation
of mass, this quantity must scale as $\rho(z)^{-2}$ so that we can write
\be
\langle n_e \rangle^\mathrm{J} (z) = \langle n_e \rangle^\mathrm{J\;base}\, \frac{\rho(z_b)^2}{\rho(z)^2}
\ee
where $\langle n_e \rangle^\mathrm{J\;base}$ is the electron number density
at the base of the jet, which is a parameter of the model. To obtain this expression,
we assume that the jet matter flows across surfaces of area $\propto \rho(z)^2$, meaning
that we consider the full jet (spine+sheath) for the mass conservation, while we consider
only the sheath for the emission. This is reasonable given that the number density
in the spine is typically extremely low~\citep{moscibrodzka13}. 

The magnetic field magnitude follows from the specification of the
magnetization parameter $\sigma$
\be
\label{eq:sigma}
\frac{B(z)^2}{4\pi} = \sigma\, m_p c^2\, \langle n_e\rangle^\mathrm{J} (z) 
\ee
where $m_p$ is the proton mass.
This means that the magnetic field will approximately
scale like $1/r$.

\begin{figure*}[htbp]
\centering
\includegraphics[width=0.4\hsize]{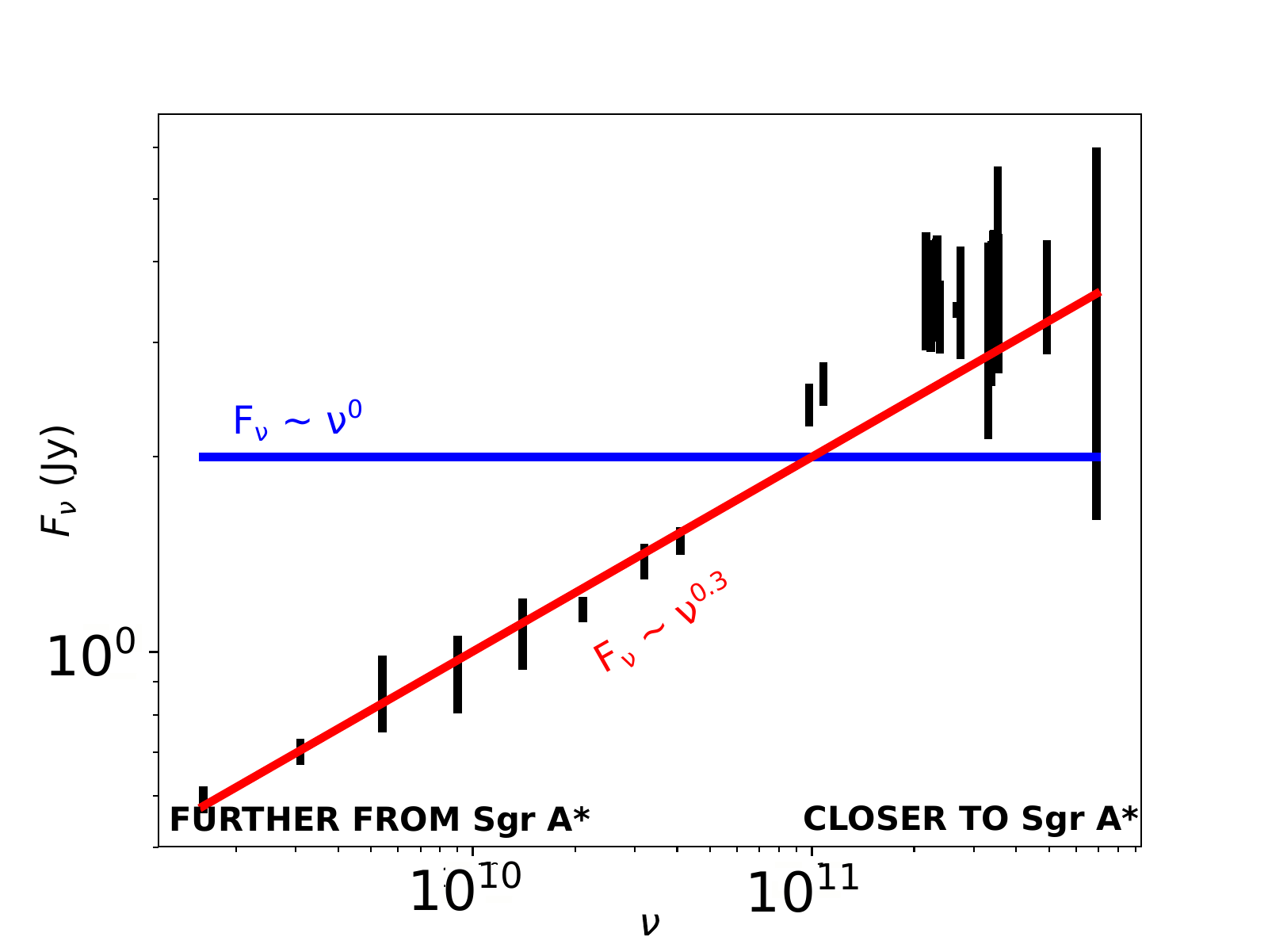}
\caption{The radio to millimeter spectrum of Sgr~A* (black error bars; see
  Fig.~\ref{fig:spectrum} for references).
  The blue horizontal line shows the spectrum produced by an isothermal jet.
  The red line shows what is needed by the data.
} 
\label{fig:radiospec}
\end{figure*}

The temperature of the electrons at the base of the jet, $T_e^\mathrm{J\;base}$,
is a parameter of the model. We are then free to prescribe the profile
of temperature with altitude. To justify our choice, let us remind that
the standard isothermal jet model of~\cite{blandford79} produces
a flat radio spectrum. For Sgr~A*, Fig.~\ref{fig:radiospec} shows
that the radio spectral data are not exactly flat: more flux is needed
at higher frequencies (closer to the black hole), rather than at lower
frequencies (further from the black hole). To obtain this behavior,
we consider a non-isothermal model such that the temperature decreases
with altitude following a power law
\be
T_e^\mathrm{J}(z) = T_e^\mathrm{J\;base} \, \left(\frac{z_b}{z}\right)^{s_T}
\ee
where the temperature slope $s_T$ is a parameter.
A value of $s_T=0$ would lead to an isothermal jet model
close to~\cite{blandford79}. We will consider values
$0 \le s_T \le 1$.

Our goal is to ray trace this model so we need to properly define the
4-velocity of the ejected gas at every points inside the jet, which will be
needed to compute redshift effects during the ray tracing. The Lorentz
factor $\Gamma_j$ being measured by the ZAMO
having 4-velocity $\mathbf{u_\mathrm{ZAMO}}$, 
the 4-velocity of the jet particles can be written 
\be
\mathbf{u_\mathrm{jet}} = \Gamma_j \left( \mathbf{u_\mathrm{ZAMO}} + \mathbf{V} \right)
\ee
where $\mathbf{V} = (V^t=0,V^r,V^\theta,V^\pp=0)$ is the jet velocity as measured by the ZAMO,
and can be seen as the usual 3-velocity of the jet. Here, it has only $2$ non-zero components due
to the axisymmetry. This jet velocity can be written easily at the external and internal
sheath surfaces (respectively defined by the angles $\theta_1$ and $\theta_2$)
\be
\mathbf{V} = V \left( \sin\theta_k \, \mathbf{e_\rho} + \cos\theta_k\, \mathbf{e}_z\right)
\ee
where $\mathbf{e_\rho}$ and $\mathbf{e_z}$ are unit vectors, $V=\sqrt{\Gamma_j^2-1}/\Gamma_j$ is the jet velocity in units of the speed of light,
and the angle $\theta_k$ can be $\theta_1$ or $\theta_2$. 
The velocity at any point inside the sheath, defined by an angle with the $\mathbf{e}_z$
direction equal to $\theta_2 < \theta < \theta_1$, can then be easily interpolated linearly
between $\theta_1$ and $\theta_2$.

The synchrotron emission and absorption coefficients for the $\kappa$-distribution
electrons are taken from~\citet{pandya16}, in their Eq.~35-41, averaged over the angle $\theta_B$ between the
magnetic field direction and the direction of emission.
The low- and high-frequency dimensionless emission coefficients,
as well as the bridging emission coefficient,  read
\bea
\mathcal{J}_\nu^\mathrm{low} &=&  X_\kappa^{1/3}\sin\theta_B\, \frac{4\pi\,\Gamma(\kappa-4/3)}{3^{7/3}\, \Gamma(\kappa-2)}, \\ \nn
\mathcal{J}_\nu^\mathrm{high} &=& X_\kappa^{-(\kappa-2)/2}\sin\theta_B \,3^{(\kappa-1)/2} \\ \nn
&& \times\frac{(\kappa-2)(\kappa-1)}{4}\, \Gamma(\kappa/4 - 1/3) \Gamma(\kappa/4+4/3), \\ \nn
j_\nu^\mathrm{\kappa-distrib} &=& \frac{\langle n_e \rangle^\mathrm{J} \,e^2\, \nu_c}{c}\,\left[ \left(\mathcal{J}_\nu^\mathrm{low}\right)^{-x_j}  + \left(\mathcal{J}_\nu^\mathrm{high}\right)^{-x_j} \right]^{-1/x_j} \\ \nn
\eea
where $X_\kappa = \nu / (\nu_c (\theta_e \kappa)^2 \sin\theta_B)$,
$\Gamma$ is the gamma function, and
$x_j=3\,\kappa^{-3/2}$. This fit is correct for $3 \leq \kappa \leq 7$. Similarly, the absorption coefficient
is given by
\bea
\mathcal{A}_\nu^\mathrm{low} &=& X_\kappa^{-2/3}\,3^{1/6}\,\frac{10}{41}\\ \nn
&& \times\frac{2 \pi}{(\theta_e \kappa)^{10/3-\kappa}}\,\frac{(\kappa-2)(\kappa-1)\kappa}{3\kappa-1}\,\Gamma(5/3) \\ \nn
&& \times  _2F_1(\kappa-1/3,\kappa+1;\kappa+2/3;-\kappa\theta_e), \\ \nn
\mathcal{A}_\nu^\mathrm{high} &=& X_\kappa^{-(1+\kappa)/2}\,\frac{\pi^{3/2}}{3}\,\frac{(\kappa-2)(\kappa-1)\kappa}{(\theta_e\kappa)^3}\\ \nn
&& \times  \left[\frac{2\,\Gamma(2+\kappa/2)}{2+\kappa} -1\right]\,
\left[ \left(\frac{3}{\kappa} \right)^{19/4} + \frac{3}{5}\right], \\ \nn
\alpha_\nu^\mathrm{\kappa-distrib} &=& \frac{\langle n_e \rangle^\mathrm{J} \,e^2}{\nu\,m_e\,c} \,\left[ \left(\mathcal{A}_\nu^\mathrm{low}\right)^{-x_\alpha}  + \left(\mathcal{A}_\nu^\mathrm{high}\right)^{-x_\alpha} \right]^{-1/x_\alpha} \\ \nn
\eea
where $x_\alpha=(-7/4+8\kappa/5)^{-43/50}$, and $_2F_1(a,b;c;z)$
is the Gauss hypergeometric function.
This function is computed by means of the implementation
of~\citet{michel08}.


\section{Spectra and images of the quiescent Sgr~A*}
\label{sec:specim}


We transport the synchrotron emission from the torus and jet
described above by means of the general-relativistic
open-source ray tracing code \gy~\footnote{See \url{http://gyoto.obspm.fr/}}~\citep{vincent11}. Null geodesics are traced backwards in coordinate time,
from the distant observer towards the black hole. The radiative transfer
equation is integrated inside the torus and jet. We have made a resolution
study to check that our choice of the technical ray-tracing parameters,
like the observer's screen resolution and field of view, ensures
a precision of $<5\%$ on the spectra. This study is briefly
summarized in Appendix~\ref{app:reso}.

Our torus+jet model is described by the set of $19$ parameters described in Table~\ref{tab:params}.
\begin{table}[htbp!]
\centering \caption{Torus+jet model best-fit parameters at inclination $i=20^\circ$.
The value of the parameter is given in bold if it is fitted.
Otherwise, the value is fixed. See text for the definition of the parameters.
For the temperatures, we also give the dimensionless electron
temperature $\Theta_e = k T_e / (m_e c^2)$. Note that the magnetic/gas
pressure ratio, defined by Eq.~\ref{eq:sigma}, is different from the
standard plasma-$\beta$ parameter. The latter is discussed 
in Eq.~\ref{eq:beta}. We note that the total mass of the
torus is of $\approx 10^{-10} \, M_\odot$.}
\begin{tabular}{l{c}r}
parameter                  &             & value                               \\
\hline
\textbf{Black hole}        &             &                  \\
mass ($M_\odot$)                       & $M$         & $4.1\times10^6$ \\
distance (kpc)                   & $D$         & $8.1~$ \\
spin                       & $a$         & $0$                                      \\
inclination ($^\circ$)               & $i$         & $20$               \\
\hline
\textbf{Torus}             &             &                 \\
angular momentum ($GM/c^3$)          & $\ell$   &  $4$                                 \\
inner radius ($GM/c^2$)              & $r_\mathrm{in}$ & $\boldsymbol{8}$      \\
polytropic index           & $k$           & $5/3$                                     \\
central density ($ \mathrm{cm}^{-3}$)           & $\langle n_e \rangle^\mathrm{T\;cen}$       &      $\boldsymbol{1.2 \times 10^9} $                          \\
  central temperature (K)            & $T_e^\mathrm{T\;cen}$   &  $\boldsymbol{7\times10^{9}}$     \\
                                     & $\Theta_e^\mathrm{T\;cen}$ & 1.2 \\
magnetization parameter             & $\sigma^\mathrm{T}$ & $0.002$                 \\
\hline
\textbf{Jet}              &                   & \\
inner opening angle ($^\circ$)       & $\theta_1$       & $20$ \\
outer opening angle ($^\circ$)       & $\theta_2$      & $\theta_1 + 3.5$ \\
jet base height ($GM/c^2$)           & $z_b$           & $2$ \\
bulk Lorentz factor        & $\Gamma_j$       & $1.15$ \\
base number density ($ \mathrm{cm}^{-3}$)  & $\langle n_e\rangle^\mathrm{J\;base}$ &  $\boldsymbol{5 \times 10^7} $\\
  base temperature (K) & $T_e^\mathrm{J\;base}$   &  $\boldsymbol{3\times 10^{10}}$\\
                       & $\Theta_e^\mathrm{J\;base}$ & 5. \\
temperature slope         & $s_T$     & $\mathbf{0.21}$ \\
  $\kappa$-distribution index & $\kappa$ &  $\mathbf{5.5}$  \\
  magnetization parameter              & $\sigma^\mathrm{J}$ & $0.01$                 \\
\end{tabular}
\label{tab:params}
\end{table}
A complete study of the parameter space would be very long in
terms of computing time, and not particularly interesting as we
may converge to solutions that are unlikely to occur in reality.
As a consequence, we prefer to rather investigate the parameter
space in a rather small region around the best-fit values of the
recent articles by~\citet{moscibrodzka13,davelaar18}.
Moreover, we will simply fix those parameters that are already
rather well constrained, or that are fully degenerate with other
parameters. In this latter case, we will choose values close to the
best-fit numerical solutions available.
The black hole mass and distance are fixed following~\citet{gravity18a}.
We consider two illustrative values for the inclination (angle between the
normal to the equatorial plane and the line of sight): either
close to face-on, $i=20^\circ$, which is in agreement
with the recent constraint from~\cite{gravity18b}, or
close to edge-on, $i=70^\circ$. As already mentioned, the spin is likely to have a small impact on the results
and is arbitrarily fixed to $a=0$. The torus constant angular momentum
scales the location of the torus center (where density and temperature
are maximum), which we decide to fix arbitrarily
at $r_\mathrm{cen} \approx 10\,GM/c^2$, leading to $\ell=4\,GM/c^3$.
The polytropic index
is fixed to $k=5/3$. The
magnetization parameters are fixed
to $\sigma^\mathrm{J}=0.01$ in the jet and $\sigma^\mathrm{T}=0.002$ in the torus.
These values were chosen by fixing the torus and jet number densities
to the best-fit values of~\citet{davelaar18} and imposing
to find similar values for the magnetic field. 
The jet inner and outer opening angles are fixed to
values similar to that describing the jet sheath of~\citet{moscibrodzka13}.
The opening angle of the jet has an overall increasing/decreasing effect
on the spectrum that would be fully degenerate with the base number density.
The jet base height is arbitrarily fixed to $2\,GM/c^2$
(coinciding with the Schwarzschild event horizon), while the bulk Lorentz
factor is fixed to $\Gamma_j=1.15$, approximately corresponding to the Keplerian velocity
at the Schwarzschild innermost stable circular orbit.
The other $7$ parameters (the central density, central temperature
and inner radius of the torus,
the base density and temperature of the jet, the temperature slope,
and the $\kappa$ index) are let free and are fitted to the spectral observations.
The temperature slope and $\kappa$ index have a strong impact on the spectrum
slope in the radio and infrared region, respectively, so that their values
can be rather easily constrained. There are thus only $5$ parameters
that we scan in detail to find our best fits.

\begin{figure*}[htbp]
\centering
\includegraphics[width=\hsize]{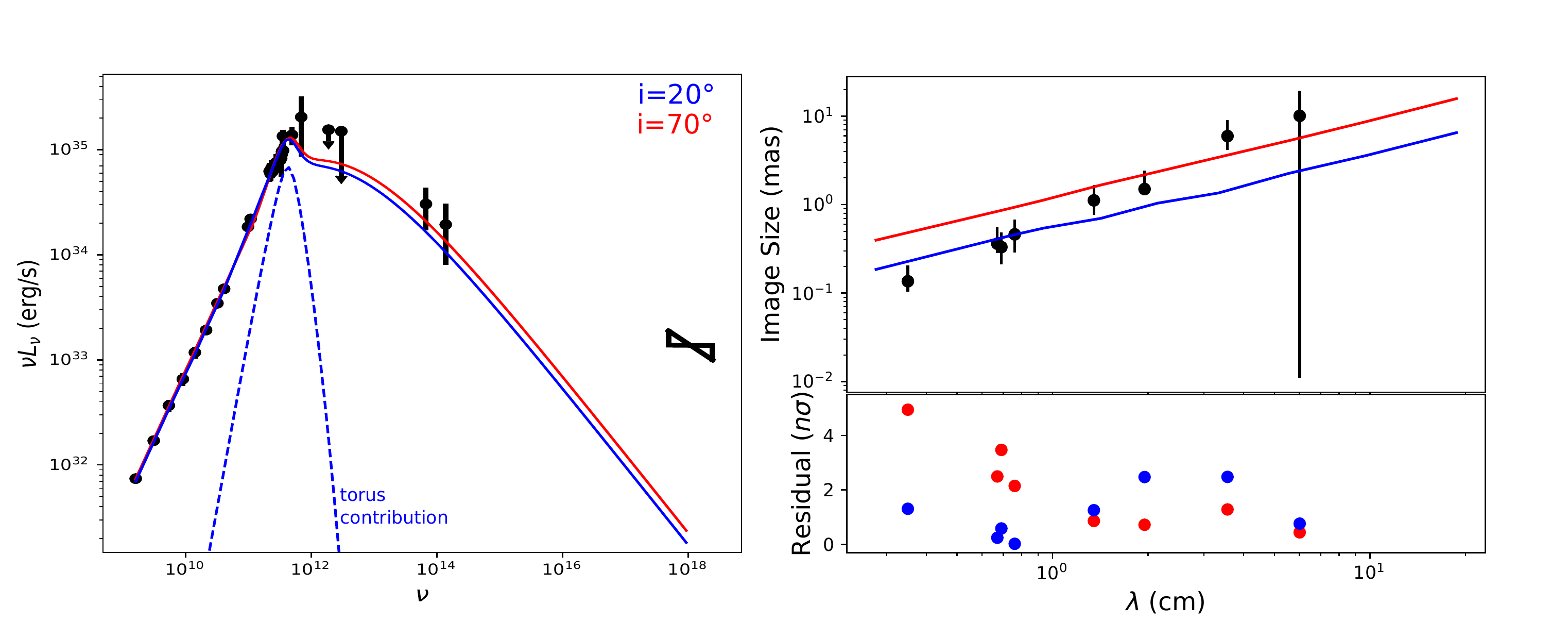}
\caption{\textbf{Left:} Best-fit torus+jet quiescent spectrum at $i=20^\circ$ (blue, $\chi^2_\mathrm{red}=0.54$, {with the torus-only contribution
    dashed; it is very similar for both inclinations so we represent only the
    $i=20^\circ$ case})
and $i=70^\circ$ (red, $\chi^2_\mathrm{red}=0.87$).
The data are taken from~\citet[][all radio data except the 4 points described just after]{bower15},~\citet[][for the 2 points around 100~GHz]{brinkerink15}, ~\citet[][for the 492~GHz point]{liu16},~\citet[][for the 690~GHz point]{marrone06},~\citet[][for the far infrared upper limits]{fellenberg18},~\citet[][for the mid infrared data]{witzel18}, and~\citet[][for the X-ray bow-tie]{baganoff01}.
Note that the X-ray data is not fitted as we do not take into account bremsstrahlung nor Comptonized emission.
\textbf{Right:} corresponding image major axis at radio wavelengths (upper panel), with data from~\citet{bower06}. The lower panel shows the residual in units of $\sigma$.} 
\label{fig:spectrum}
\end{figure*}

As we will see in Fig.~\ref{fig:spectrum}, the torus has a non-negligible
impact on the spectrum only in the millimeter band. As a consequence, we have
divided our parameter space scanning into two easier sub-tasks: first we
fit a pure jet (with no torus) to the radio and infrared spectral data
(removing the millimeter peak).
Then we fix the jet parameters at their values at the best fit of this search
and fit the torus parameters by fitting the full spectrum (from radio
to infrared). This allows to decrease
the dimensions of the parameter space and thus save computing time.
We thus considered one 2D grid (density, temperature) for the jet, and one
separate 3D grid (density, temperature, inner radius) for the torus.
We note that we fit our model to the spectral data only. The validity of the
constraints on the radio image size and infrared spectral index are then
checked a posteriori.

The left panel of Fig.~\ref{fig:spectrum} shows the best-fit spectra obtained
for the two values of inclination considered here, $i=20^\circ$
and $i=70^\circ$.
The fits are extremely good, with reduced chi-squared
$\approx 1$ for both inclinations. The best-fit parameters 
are listed in Table~\ref{tab:params} for the $i=20^\circ$ case.
Those of the $i=70^\circ$ case are
the same except for the following: $r_\mathrm{in}=6\,GM/c^2$,  $\langle n_e \rangle^\mathrm{T\;cen} = 8.7\times 10^8 \, \mathrm{cm}^{-3}$, $T_e^\mathrm{T\;cen} = 6\times 10^9$~K,
$\langle n_e\rangle^\mathrm{J\;base} = 7.5\times10^7 \, \mathrm{cm}^{-3}$.
It is interesting to compare the jet-base and torus-center values of the
number density, temperature and magnetic field to the best-fit GRMHD
simulation of~\citet{davelaar18}. We checked that our best-fit values
are within a maximum factor of $\approx 2.5$ from that of these
authors. This is a rather strong argument in favor of the robustness
of both methods. It also shows that our much simpler analytic model
does capture the essential aspects of the physics at play.
We also note the striking similarity between our best-fit non-thermal spectra
and that of~\citet{yuan03}, developed for the very different context
of RIAFs~\citep[see e.g. the recently updated Fig.~19 of][]{witzel18}.
We consider that such comparisons are strong arguments in favor
of the robustness of the simulations of Sgr~A* accretion flow.

At this point, it is interesting to compute what is our best-fit
plasma $\beta$ parameter,
as defined in the standard way of the ratio
between the thermal to magnetic pressure ratio. We find
\be
\label{eq:beta}
\beta = 8\pi \, \frac{\langle n_e \rangle \, k \, T_e}{B^2} \approx 0.6
\ee
which is valid both at the center of the torus and at the base of the jet.
Our plasma is thus close to be fully magnetized (i.e. to $\beta=1$).
This value is comparable to the inner disk $\beta$ of~\citet{ressler17},
as reported in their Fig.~1, lower-left panel.

Based on the recent detailed analysis of the infrared statistical
properties of Sgr~A* by~\citet{witzel18}, we can also discuss the
value of our predicted infrared spectral index. Here, we define this
index as the factor $\alpha$ such that the specific infrared flux
follows $F_\nu \propto \nu^{-\alpha}$. This parameter is easily
related to the $\kappa$ index of our electron distribution through
$\alpha = (\kappa-2)/2$. With our best-fit value of $\kappa=5.5$,
the predicted spectral index of our model is thus $\alpha^\mathrm{predict}=1.75$.
The (dereddened) $2.2\,\mu$m luminosity of our $i=20^\circ$ best-fit model
reaches $\nu L_\nu = 1.06\times 10^{34}$~erg/s.
Using Table~6 of~\citet{witzel18}, {which gives the relation
between dereddened and non-dereddened fluxes of Sgr~A*},
this translates to a non-dereddened flux of order $0.1$~mJy.
Using now Fig.~17 of~\citet{witzel18}, {which gives the K-band
  spectral index as a function of the non-dereddened flux},
this translates to a spectral
index of the order of $\alpha^\mathrm{obs}=1.8$, so very close to our
predicted value. The slightly higher flux value at $i=70^\circ$ leads
to similar conclusions. Thus, our best-fit models are coherent with
the quiescent constraints on Sgr~A*'s spectral index.

The third and final observable that we can use are the intrinsic radio
sizes of Sgr~A*. The right panel of Fig.~\ref{fig:spectrum} shows the
predicted major axis size of our best-fit models for both inclinations,
compared to the data of~\citet{bower06}. The major axis of the image
is computed from the image central moments. We briefly remind this
formalism in Appendix~\ref{app:moments}.
The right panel of Fig.~\ref{fig:spectrum} shows that our $i=20^\circ$
best-fit model is always at $<2.5\sigma$ from the $0.35$ to $6$~cm data
of~\citet{bower06}, which gives a reasonable agreement over this range.
However, our model predicts that the size evolves like
$\lambda^\gamma$, where $\gamma \approx 0.8$, which is too shallow
with respect to the constraint of~\citet{bower06}, who find
that $\gamma^\mathrm{obs} \approx 1.6$. This leads our model to
overpredicting the size of the image at lower wavelengths, as
we will see below. We note that the centimeter-size behavior of our model
is very similar to that depicted in Fig.~9 of~\citet{davelaar18},
again showing the ability of our simple description to lead
to the same conclusions as the most sophisticated GRMHD simulations to date.
{Let us stress that the slope of the curve on the righ panel of
  Fig.~\ref{fig:spectrum} is only weakly dependent on the jet parameters.
  \citet{davelaar18} and~\citet{chael18} have shown that the size of the cm
  emitting region is sensitive to the electron distribution function, so that
  the shallow slope that we get might be linked to our choice of purely
$\kappa$-distribution electrons.}

\begin{figure*}[htbp]
\centering
\includegraphics[width=\hsize]{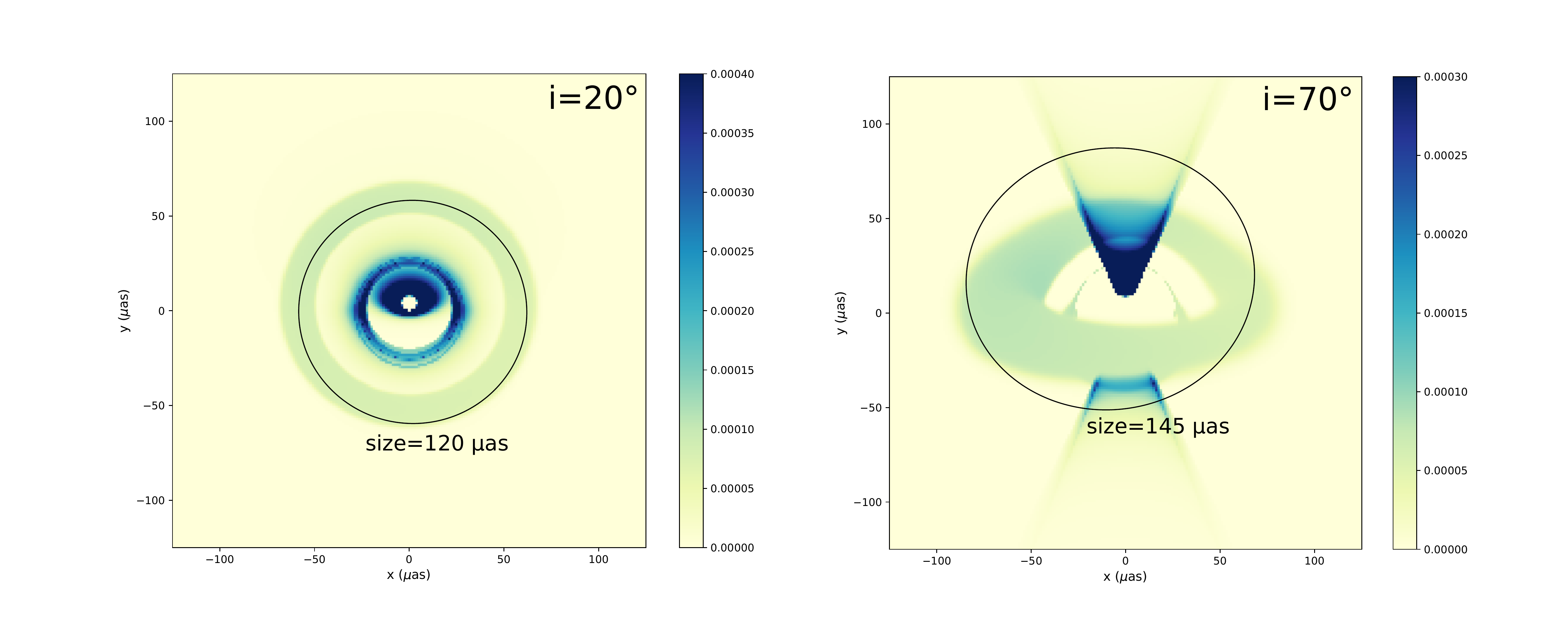}
\caption{Best-fit torus+jet image at $1.3$~mm, as seen at inclination $i=20^\circ$ (left)
  and $i=70^\circ$ (right). The color bar is different for the two images to optimize the readability of each panel. It gives the value of the specific intensity
  in cgs units. The color hue is somewhat saturated to make it easier to see the fainter torus. As a consequence, the maximum value of the specific intensity is somewhat higher than the highest number of the color bars: $0.0015$ for the
  left panel, and $0.001$ for the right panel, in cgs units. The black ellipses are obtained by
  deriving the central moments of the images (see text for details) and give an estimate of the size of the emitting region, which is written explicitly in
each panel.} 
\label{fig:image}
\end{figure*}

\citet{doeleman08} have given a constraint of the intrinsic diameter
of Sgr~A* at $1.3$~mm of $37^{+16}_{-10}\,\mu$as ($3\sigma$),
based on a Gaussian fit.
It is thus
particularly interesting to examine the prediction of our model
at this specific EHT wavelength.
Fig.~\ref{fig:image} shows the $1.3$~mm best-fit image of our model
for both inclinations.
It shows that our predicted $1.3$~mm size (as computed from image moments)
is larger by a factor of
$\approx 3$ at $i=20^\circ$ and $\approx 4$ at $i=70^\circ$,
as compared to the~\citet{doeleman08} constraint. This
is mainly due to the presence of the faint extended torus,
while our images also show prominent features
at the $\lesssim 40\,\mu$as scale.
The time-evolving GRMHD model of~\citet{davelaar18}
leads, here again, to very similar results.
The size millimeter constraint reported above
is valid assuming a circular Gaussian model for the source.
A thick-ring model leads to an outer diameter intrinsic source size
of $\approx 80\,\mu$as, so a factor $\approx 1.5$ smaller
than our face-on prediction.
The constraint of~\citet{doeleman08} is thus only the first word on a nascent
topic. In particular, this constraint is only valid in the projected
direction of the baseline on sky, so that a complex geometry (like we have
here with a thick disk and a jet),
with an intrinsic size varying a lot with the angle on sky,
might be too broadly described by this single number only.
Therefore, we consider that our $1.3$~mm flux repartition
is in reasonable agreement with the data.
It is likely that the near-future EHT data will allow to
constrain more precisely the geometry of the inner accretion flow,
allowing to further refine the modeling part.

\begin{figure*}[htbp]
\centering
\includegraphics[width=\hsize]{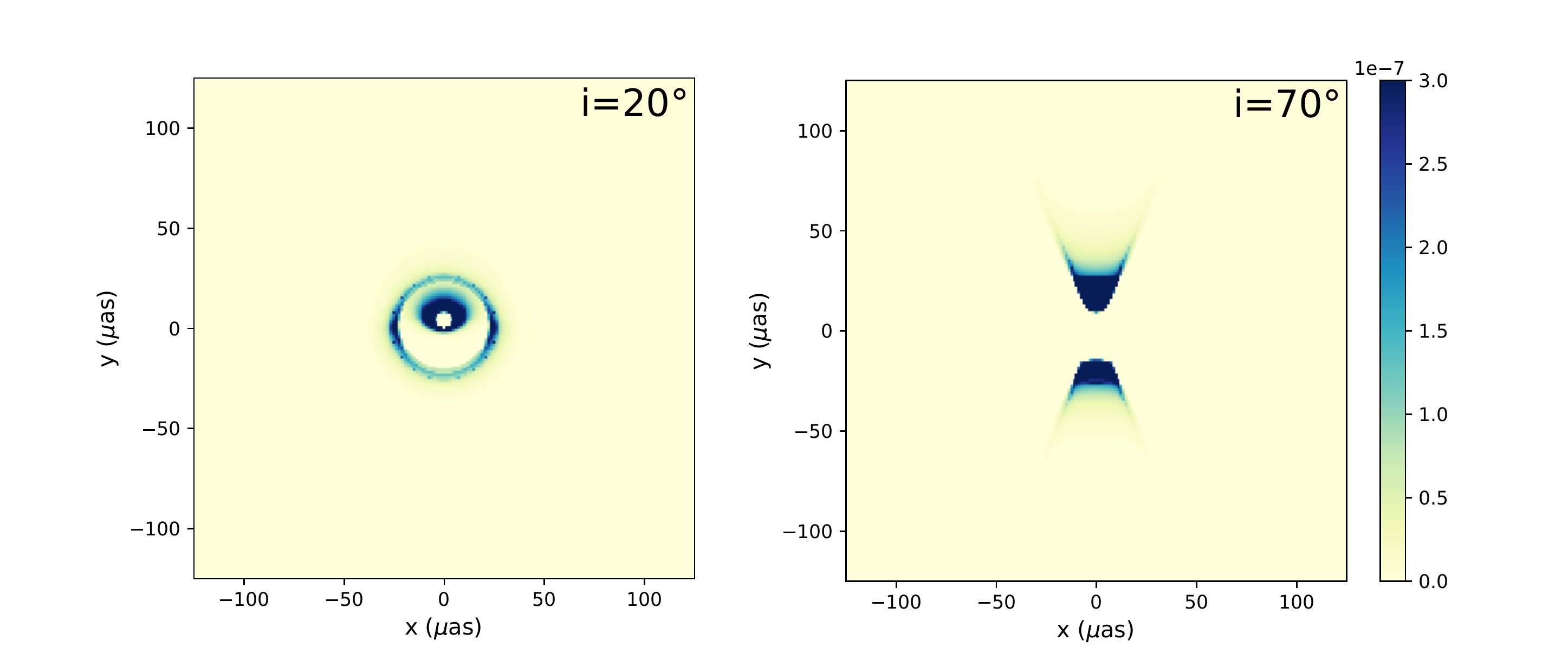}
\caption{Best-fit torus+jet image at $2.2\,\mu$m, as seen at inclination $i=20^\circ$ (left)
  and $i=70^\circ$ (right). The color bar is common to both panels
  and gives the values of specific intensity in cgs units. As in
  Fig.~\ref{fig:image}, the color hue is somewhat saturated for better
  visualization.} 
\label{fig:gravimage}
\end{figure*}

Fig.~\ref{fig:image} shows that the $1.3$~mm image is due to
a mix of contributions due to the torus and the jet. At radio
wavelengths ($< 10^{11}$~Hz), the jet completely dominates the
spectrum, as emission primarily comes from large scales.
Our model is also fully dominated by the jet for near infrared
frequencies and above, as illustrated in Fig.~\ref{fig:gravimage},
which shows the best-fit $2.2\,\mu$m images at both inclinations.
This feature is in reasonable agreement with the near infrared
images of~\citet{davelaar18}. However, this disagrees with the
results of~\citet{ressler17} who find that the disk dominates
at all frequencies above the millimeter peak. This difference is
certainly due to the different electron temperatures in the various
models. In particular, \citet{ressler17} report hot spots
of high electron temperature in the disk, that are obviously not
present in our simple setup. These hot spots lead to a high
near-infrared flux, which would not agree with the faintest quiescent
level of Sgr~A*.

\begin{figure*}[htbp]
\centering
\includegraphics[width=0.4\hsize]{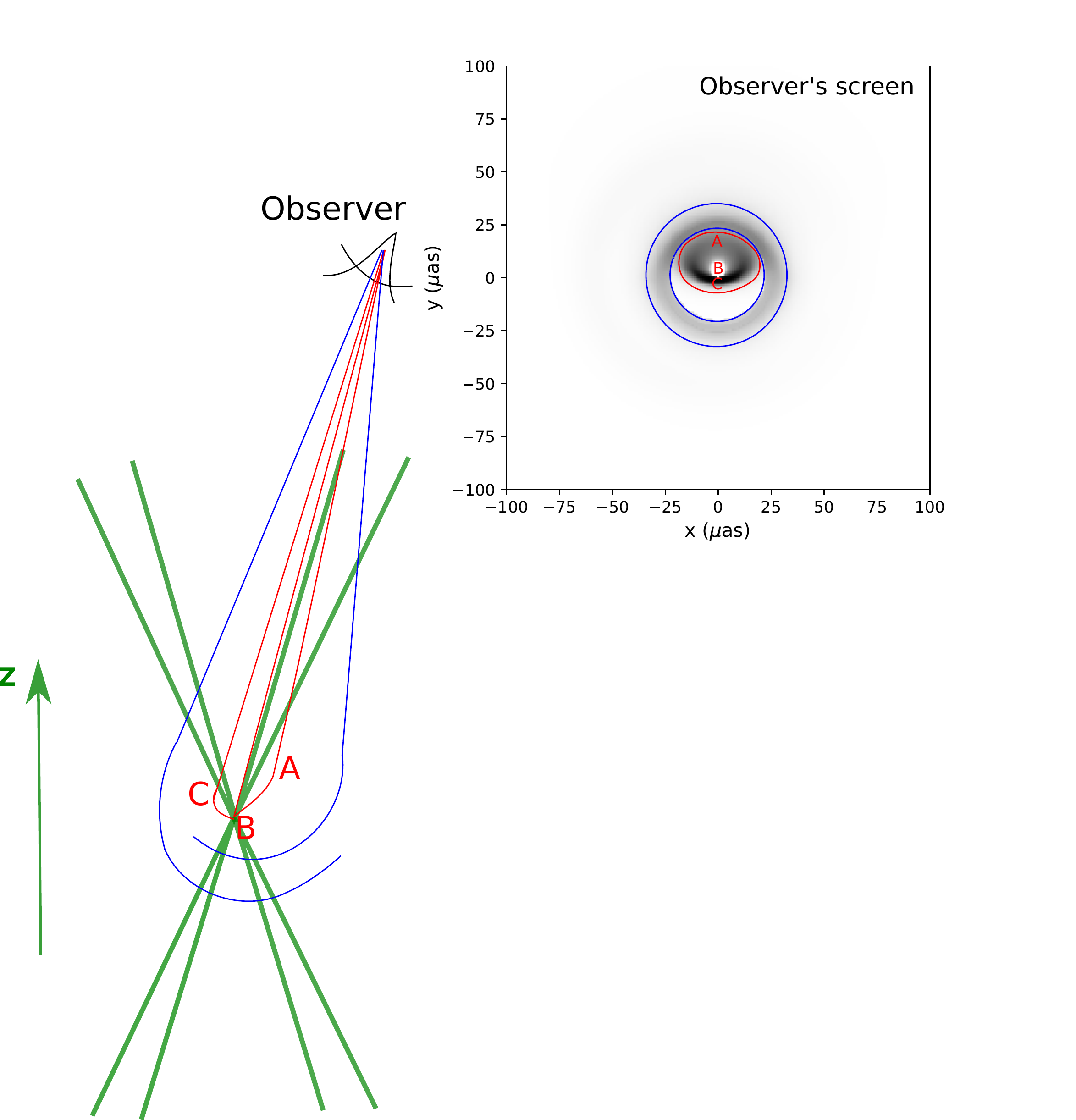}
\caption{The green contours show a scheme of the jet sheath
  on both parts ($z>0$ and $z<0$) of the black hole.
  The observer is at $i=20^\circ$ at the top of the figure,
  with the observation screen represented at the top right.
  It is identical to the left panel of
  Fig.~\ref{fig:gravimage}. Two regions are highlighted in red and blue.
  The red part is the primary image of the regions close to the $z>0$ base
  of the jet. Three examples of geodesics, ray-traced backwards in time
  from the observer's screen,
  A, B, and C, are represented on the scheme, all of them ending their
  trajectory inside the black hole.
  Their end points on the observer's screen are labeled.
  Geodesic C carries more flux, because it has visited regions
  very close to the base of the jet. While geodesic B, falling (backwards)
  straight
  into the black hole, carries no flux.
  The toroidal blue region on the observer's screen
  is the secondary image of the jet, due to photons
  that are strongly bent in the regions close to the $z<0$ base of the jet,
  before reaching the far-away observer. Two such strongly-bent geodesics
  are depicted in blue on the scheme. This structure can thus be seen as the Einstein ring
of the $z<0$ base of the jet.} 
\label{fig:imjet}
\end{figure*}

Although the right panel of Fig.~\ref{fig:gravimage}, showing the edge-on ray-traced
image of a jet, is easy to interpret,
it is likely that the left panel, showing the same scenery from a
face-on view, is not so. 
Fig.~\ref{fig:imjet} tries to explain this image,
showing that the annular structure
is actually the Einstein ring of the $z<0$ base
of the jet.

\section{Reconstructing synthetic data with the EHT array}
\label{sec:ehtimaging}
An important question to ask is whether salient features of Sgr~A* near-horizon emission region, that we are parametrizing with analytic geometric models, could actually be observed by an instrument such as the EHT. The question concerns not only the instrument resolution, but also inherent limitations of the imaging from sparsely sampled Fourier domain data  and utilizing a~strongly inhomogeneous array of telescopes, both being traits of VLBI in general and EHT in particular. One of the limitations is a low dynamic range of VLBI synthesis images, see, e.g., \cite{braun2013}. For a multicomponent source this could result in the inability of the EHT observations to reliably detect a~weaker-flux component, such as a faint torus in the~presence of a~bright jet.

We investigate this issue by generating synthetic EHT observations of the images shown in Fig. \ref{fig:image} and subsequently attempting to reconstruct the images from sparsely sampled data. Synthetic observations and image reconstructions are generated using the~freely available \texttt{eht-imaging} library~\footnote{\url{https://github.com/achael/eht-imaging}}. A Maximum Entropy Method (MEM), implemented in \texttt{eht-imaging}, was used for the image reconstruction, see \cite{chael2016}. The simulated observations fold in characteristic sensitivities of the EHT telescopes, and effects such as thermal noise contamination, rapid atmospheric phase variation, and dependence of sensitivity on source elevation. The EHT 2017 array was used, with optimal coverage in the Fourier domain. A static source model (i.e., single image) was assumed, which is a~big simplification, as time variability on timescales as short as minutes is expected for Sgr~A*. No mitigation of scattering, subdominant for $1.3$~mm wavelength, was employed.

\begin{figure*}[htbp]
\centering
\includegraphics[width=0.325\hsize]{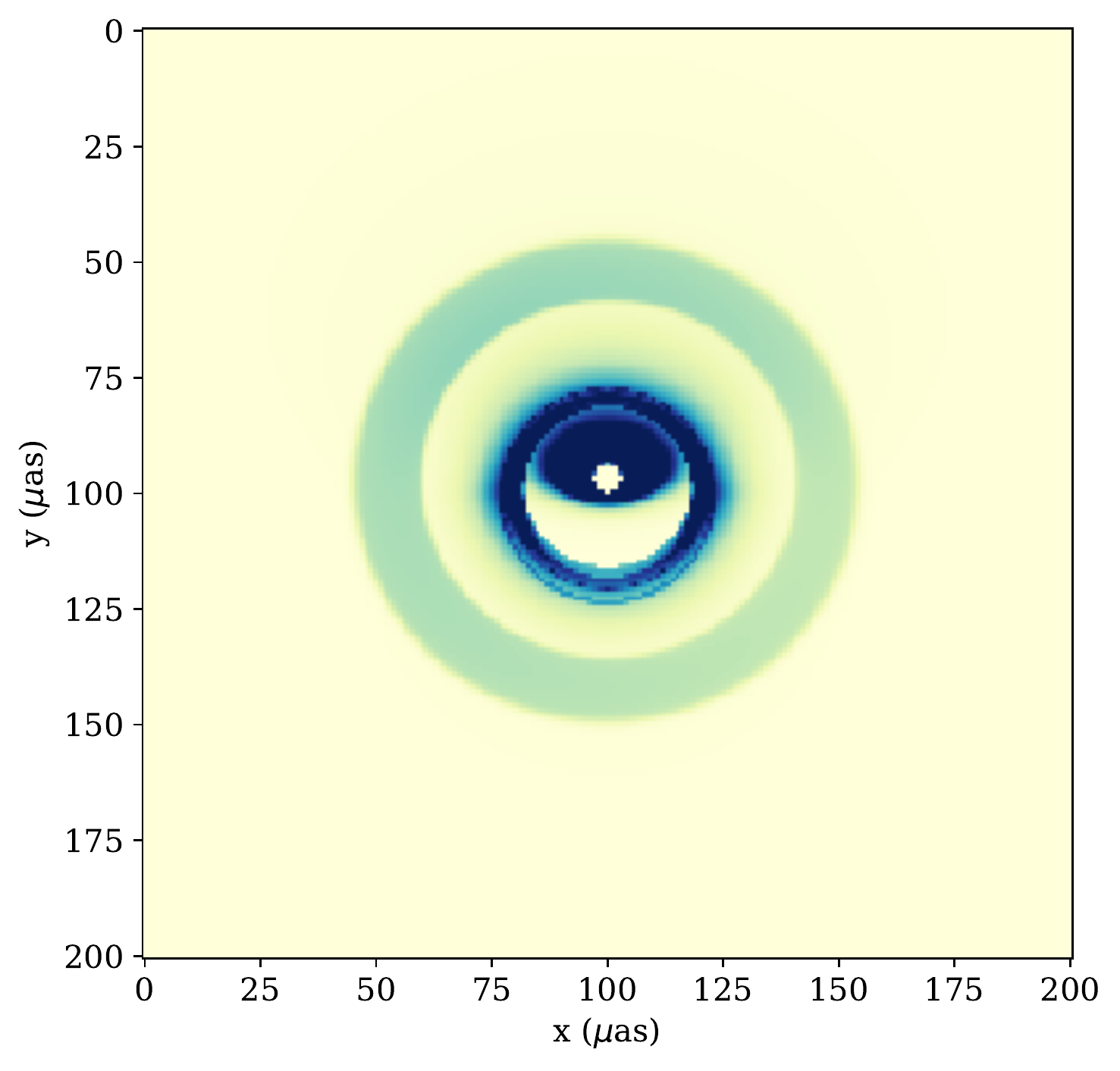}
\includegraphics[width=0.325\hsize]{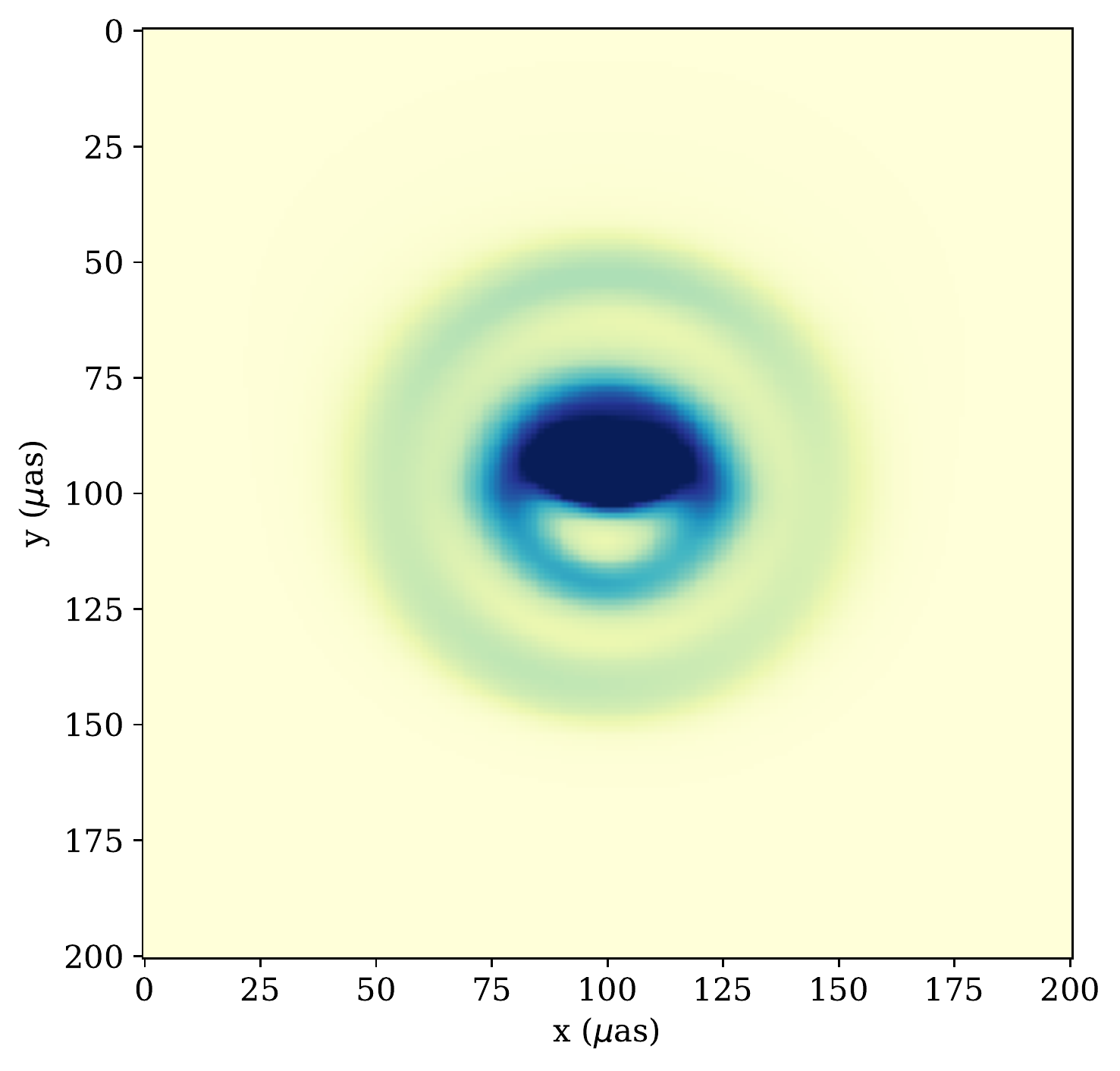}
\includegraphics[width=0.325\hsize]{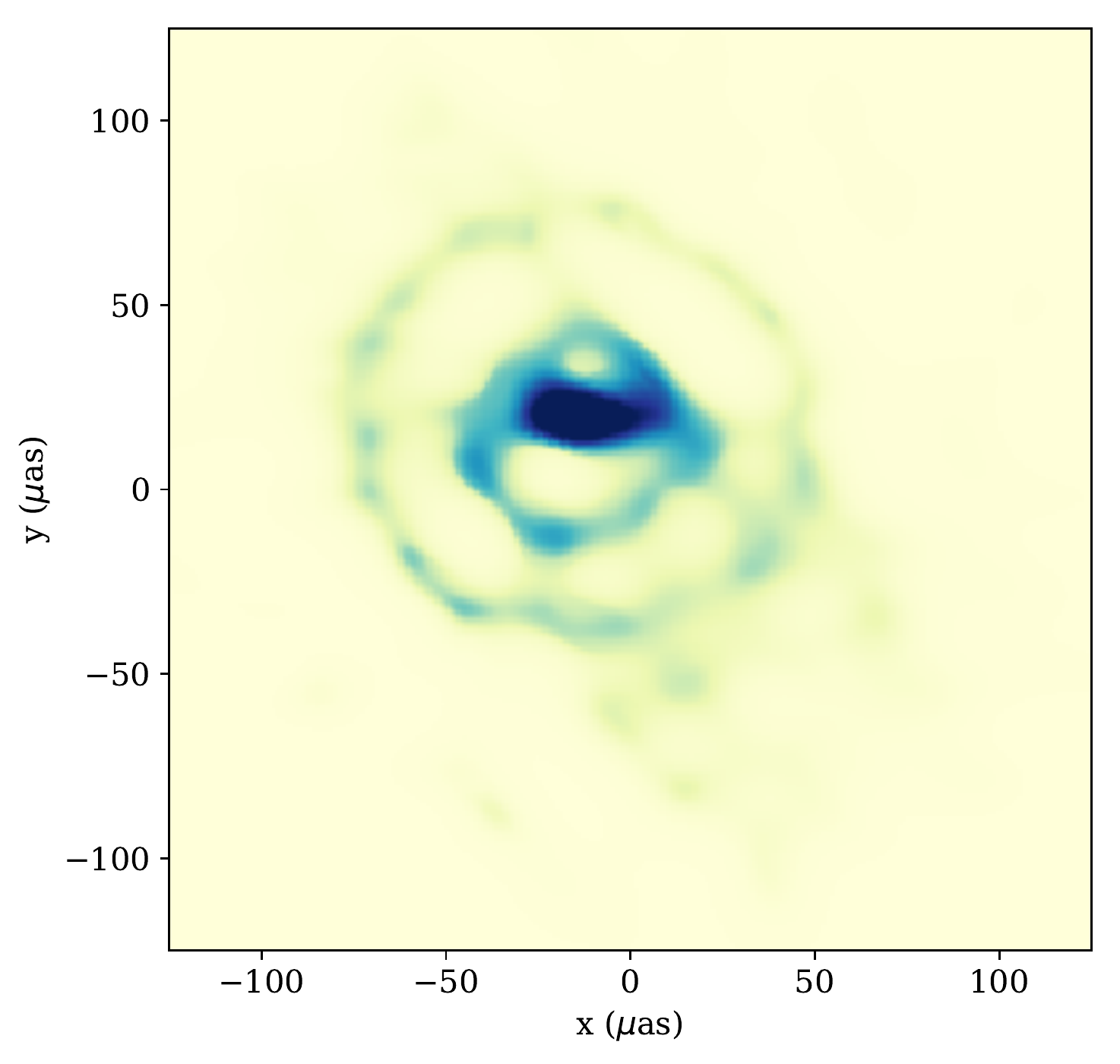}
\includegraphics[width=0.325\hsize]{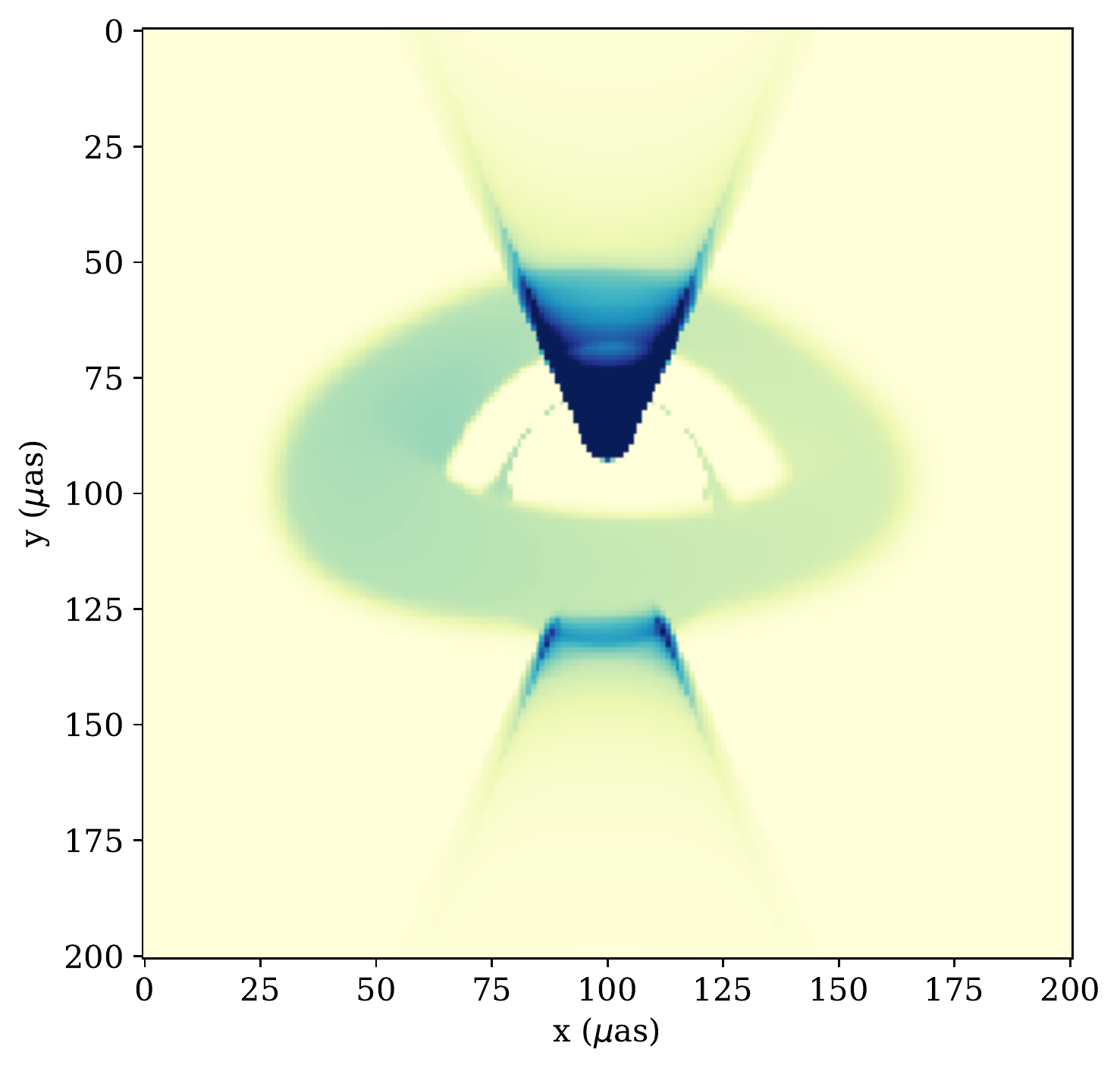}
\includegraphics[width=0.325\hsize]{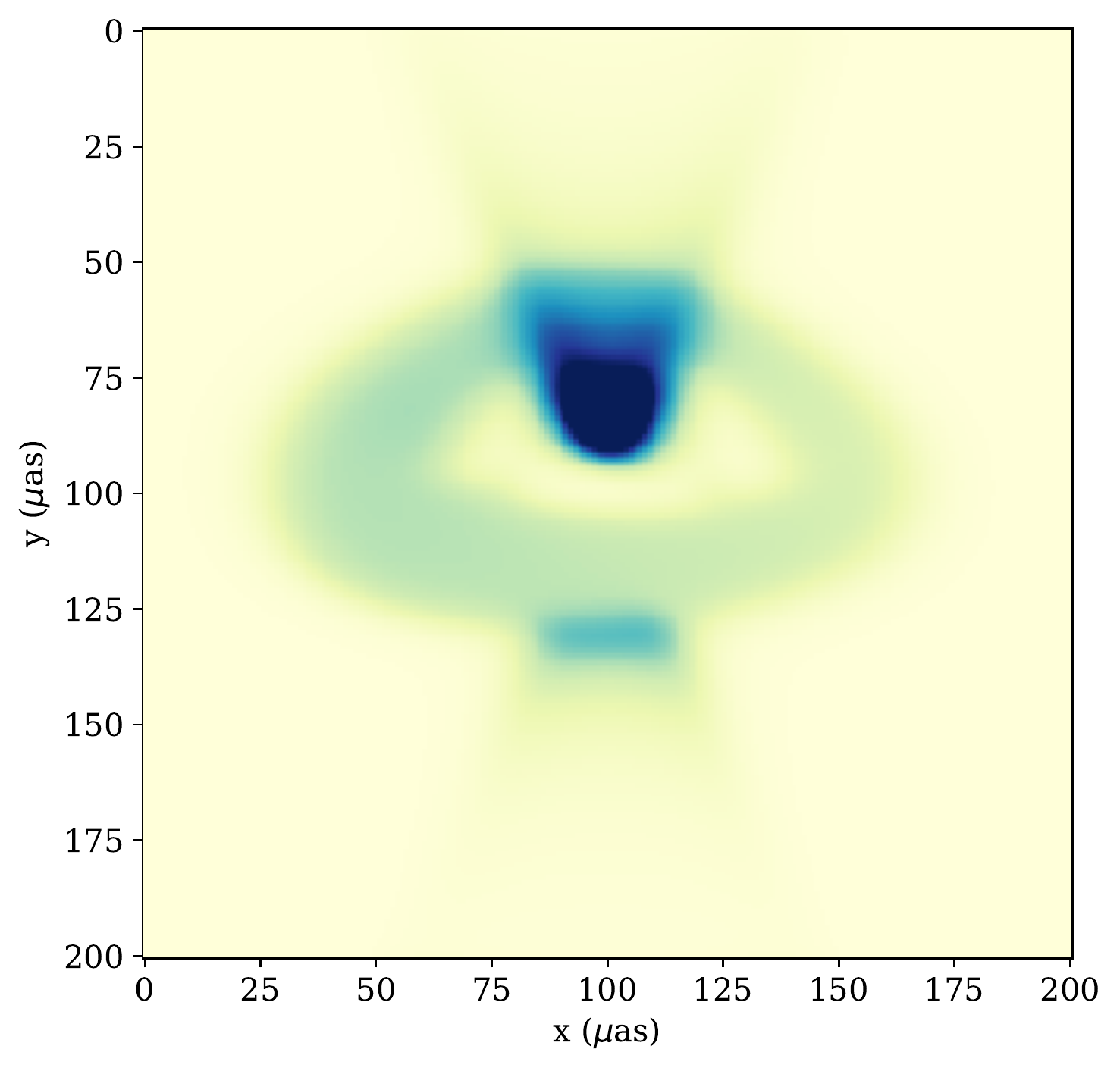}
\includegraphics[width=0.325\hsize]{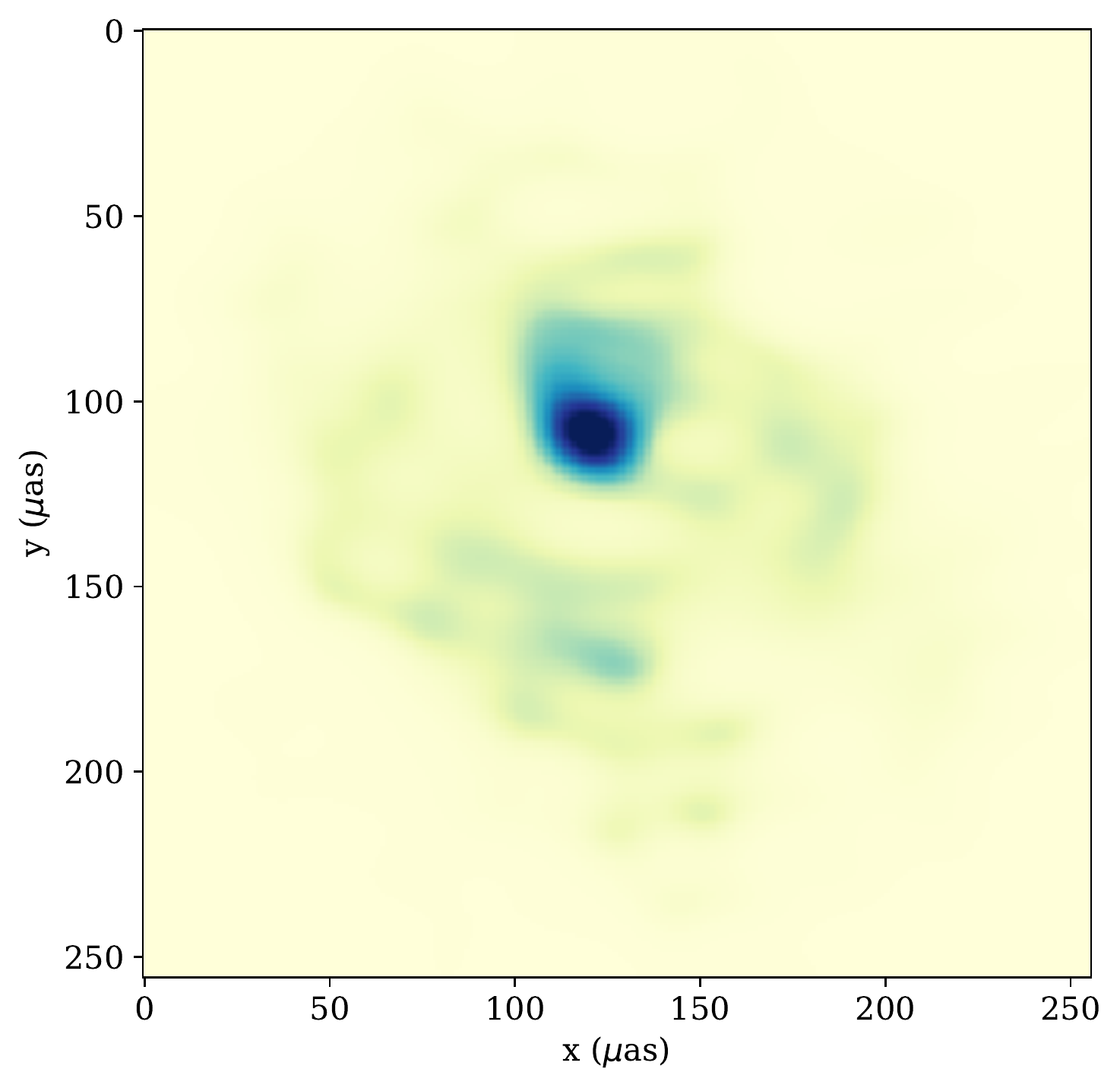}

\caption{{Example of model images reconstruction with the synthetic EHT array. For display purposes colormaps saturates at 0.001 Jy/px. Left column: original synthetic images generated with our model (same as Fig.~\ref{fig:image}). Middle column: model images as processed by the interstellar scattering screen \citep{johnson2018}. Right column:  MEM reconstruction of images, observed by a synthetic EHT 2017 array \citep{chael2016}. Fitting elliptical Gaussian component to images we find characteristic size [major axis, minor axis] of [97,72] $\mu$as for the top row image and [100,89] $\mu$as for the bottom row image.  } }
\label{fig:eht-imaging}
\end{figure*}

EHT reconstruction results are shown in Fig.~\ref{fig:eht-imaging}. They show that the salient features of the models persist in the reconstructed images. In particular, a wide region of weak emission, corresponding to the faint torus, is present in the reconstructed images.
This successful reconstruction of the model images allows us to hope that similar features of the realistic Sgr~A*
accretion/ejection flow could be successfully revealed in EHT images.
If so, simple geometric models such as ours could help interpreting
future data and extracting relevant parameters, such as, in the present case,
the inclination angle. Fig.~\ref{fig:eht-imaging} indeed shows
that the reconstructed image is clearly dependent on this important parameter.

\section{Conclusion and perspectives}
\label{sec:conc}

We present here a simple analytic model of the quiescent-state
emission of Sgr~A*,
made of the combination of a compact torus and a large-scale
jet sheath. Our model allows to fit very well the multi-wavelength
spectral data of Sgr~A*, as illustrated in the left
panel of Fig.~\ref{fig:spectrum}.
The size of the radio/millimeter emitting region is in reasonable agreement with
observed constraints, as illustrated in the right
panel of Fig.~\ref{fig:spectrum} and in the discussion
accompanying Fig.~\ref{fig:image}.
Our Fig.~\ref{fig:eht-imaging} demonstrates
that salient disk/jet features of our model images persist
when synthetic data are 'observed' and reconstructed
using a~numerical model of the EHT array, and that these features
are sensitive to inclination.

It is interesting that our model, inspired by the recent
work of~\citet{davelaar18}, leads to best-fit parameters
very close to that found in the GRMHD simulations of these
authors. We believe that this is a nice illustration of the
interest of simple analytic models: they are able to reproduce
the outputs of costly numerical simulations.
It is also interesting that our spectral prediction is
indistinguishable from the predictions of~\citet{yuan03},
who use a different analytic description of the surroundings
of Sgr~A*. This is a good argument that the theoretical descriptions
of Sgr~A* are robust in their predictions.

We consider that our model is a practical testbed
to study various aspects of the physics of Sgr~A*.
We hope that this model can be useful for other authors,
and describe all steps necessary to reproduce
our results in Appendix~\ref{app:gyoto}.
In the near future, we aim at using this model to analyze and interpret
the data from GRAVITY, EHT and possibly other millimeter-range
VLBI observations.

\appendix

\section{Resolution study}
\label{app:reso}

The emitting part of the jet in the radio range can extend
to large distances, imposing to consider a large field of view
for the ray tracing computation. Here, we investigate the
resolution of the \gy screen (i.e. the number of pixels
along one dimension, labeled $N$) needed to obtain a precise
value of the observed flux. We want to determine the optimal
pair of field of view $F$ and resolution $N$.
\begin{figure*}[htbp]
\centering
\includegraphics[width=0.5\hsize]{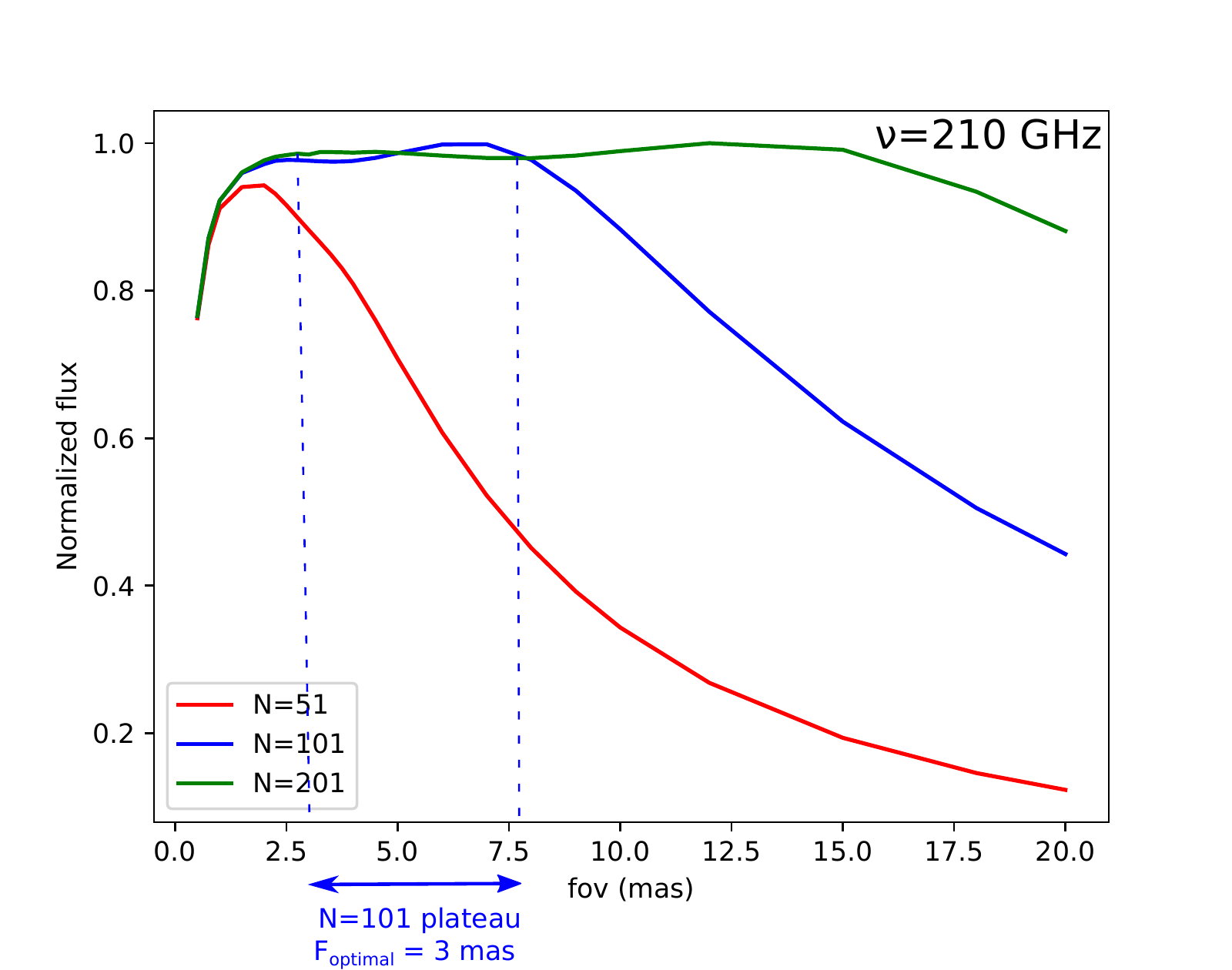}
\caption{Evolution of the normalized ray-traced flux
at $\nu = 210$~GHz with increasing field of view for $3$ values of screen resolution, $N=51$ (red), $N=101$ (blue), and $N=201$ (green). The optimal value of field of view for a given
resolution is the beginning of the plateau-phase of the curve (see text for details). The plateaus of
the $N=101$ and $N=201$ curves are identical to within $0.1 \%$,
which shows that $N=101$ is the smallest resolution ensuring accurate
flux values, associated to $F=3$~mas.
}
\label{fig:resostudy}
\end{figure*}
To do so, we study the evolution of the normalized flux with
the field of view, for various resolutions and for
a set of wavelengths.
The overall behavior of these curves is easy to understand. 
For a given resolution, if the field of view is too small, 
the predicted flux is also too small because a portion of the
emitting region leaks out of the field of view. If the field of
view is too big, the predicted flux will also be too small, 
because the emitting region is diluted (at the limit of
a field of view $4\pi$ steradian, the emitting region would
be so small that the image would be completely black, leading
to zero flux). Thus the curve showing the evolution of the
flux as a function of the field of view first increases
with the field of view, then stabilizes to form a plateau, and finally
decreases.
We select our $(N,F)$ pairs by imposing
that the plateaus of the curves corresponding to $N$ and to $2N$
are equal to within $<5\%$. For the minimal $N$ satisfying
this condition, we choose the smallest value of $F$ within the plateau.
A smaller $F$ will lead to a smaller computing time (because the region to
trace is smaller), so that this is the optimal choice in terms of
both precision and computing time.
Fig.~\ref{fig:resostudy} illustrates this procedure for the particular case
of $\nu = 210$~GHz.
This figure shows
that the plateau fluxes corresponding to the $N=101$ and $N=201$ curves
are equal
to within $0.1~\%$, while the $N=51$ plateau is $6\%$ off and thus rejected.
Table~\ref{tab:NFpairs} gives the various $(N,F)$ used in this article
as a function
of the observed frequency.

\begin{table}[htbp!]
  \centering \caption{Pairs of resolution $N$ and corresponding
    field of view $F$ used for the $i=20^\circ$ simulations
    of Section~\ref{sec:specim}.}
\begin{tabular}{cc}
$\nu$ (Hz)                  &    $(N,F \mathrm{(mas)})$                              \\
\hline
  $1.6\times 10^9$               &     $(101,30)$             \\
  $3.1\times 10^9$               &     $(101,15)$             \\
  $5.4\times 10^9$               &     $(101,10)$             \\
  $9\times 10^9$               &     $(101,5)$             \\
  $1.4\times 10^{10}$               &     $(101,5)$             \\
  $2.1\times 10^{10}$               &     $(101,3)$             \\
  $3.2\times 10^{10}$               &     $(101,3)$             \\
  $4.09\times 10^{10}$               &     $(101,2.5)$             \\
  $1.08\times 10^{11}$               &     $(201,1)$             \\
  $> 1.1\times 10^{11}$ &    $(201,0.5)$ \\
\end{tabular}
\label{tab:NFpairs}
\end{table}

\section{Image moments}
\label{app:moments}

Let $I(x,y)$ be a 2D image labeled by a Cartesian
grid $(x,y)$. The central moment of order $p+q$ of
image $I$ is the quantity
\be
\mu_{pq} = \sum_x \sum_y (x-\bar{x})^p (y-\bar{y})^q \, I(x,y)
\ee
where $(\bar{x},\bar{y})$ is the centroid of the $I(x,y)$ distribution,
i.e.
\be
\bar{x} = \frac{\sum_x \sum_y x \, I(x,y)}{\sum_x \sum_y I(x,y)}
\ee
and similarly for $\bar{y}$.

The major axis of the best-fitting ellipse adjusted to the distribution
of $I(x,y)$ in the image is then given by~\citep{birchfield18}
\be
L = 2 \sqrt{2\,\frac{\mu_{20} + \mu_{02} + \sqrt{(\mu_{20}-\mu_{02})^2 + 4\mu_{11}^2}}{\mu_{00}}}
\ee
while the orientation of the ellipse with respect to the
Cartesian grid $(x,y)$ is
\be
\tan 2\theta = \frac{2 \mu_{11}}{\mu_{20}-\mu_{02}}.
\ee

The size-fitting ellipses of Fig.~\ref{fig:image} are computed
using these formulas, as implemented in the \texttt{cv2} Python package.

\section{Using \gy to generate spectra and images}
\label{app:gyoto}

{The code developped for this paper is part of
\gy 1.3.1~\citep[][also available at
\url{https://github.com/gyoto/Gyoto/tree/1.3.1}]{paumard19}.
\gy is packaged for Debian
GNU/Linux and its derivatives including Ubuntu and this version will
be part of the next version of these operating systems to be released
in 2019}. The installation steps are detailed in the file
INSTALL.Gyoto.md (skipping section 0: the pre-compiled versions of \gy
do not contain the very recent new developments presented in this
article).

The input file Gyoto/doc/examples/example-jet.xml gives the jet-only
best-fit model for the $i=20^\circ$ case discussed in Section~\ref{sec:specim}.
The file Gyoto/doc/examples/example-torusjet.xml gives the torus+jet
best-fit model, i.e. the model used to generate the face-on spectrum
and image of Fig.~\ref{fig:spectrum} and~\ref{fig:image}. The xml
files provided have parameters such that they allow an accurate computation
of the spectrum in the $10^{11}$ to $10^{18}$ Hz range. Lower frequencies
need higher resolution and longer computing time, see Appendix~\ref{app:reso}.

The Python scripts Gyoto/doc/examples/plot-Spectrum.py and
Gyoto/doc/examples/plot-Image.py allow to straightforwardly
generate spectra (together with the latest observed data)
and images (together with the best-fitting image-moment ellipse),
just as in our Fig.~\ref{fig:spectrum} and~\ref{fig:image}.

We thus provide all the software needed to obtain the results presented
in this article.

Interested people are very welcome to contact the \gy developers
at \url{frederic.vincent@obspm.fr}, \url{thibaut.paumard@obspm.fr}
to get help.

\section*{Acknowledgements}
{FHV acknowledges fruitful inputs from T. Bronzwaer, J. Davelaar and G. Witzel.
  FHV acknowledges many interesting discussions at the \textit{Central
    Arcsecond} conference in Ringberg (Nov. 2018), and would like to thank
  T. Do, H. Falcke, S. von Fellenberg, D. Wang, and the organizers of the conference. FHV acknowledges interesting email exchanges with F. Yuan.
  MAA acknowledges the Polish NCN grant 2015/19/B/ST9/01099 and the
Czech Science Foundation grant No. 17-16287S which supported
his visits to Paris Observatory and to Harvard University; Harvard's
Black Hole Initiative support is also acknowledged.
  AAZ has been supported in part by the Polish National Science Centre
grants 2013/10/M/ST9/00729 and 2015/18/A/ST9/00746.
  }
%
%
\bibliography{torusjet}
\bibliographystyle{aa}
\end{document}